\documentclass{article}
\usepackage{sgamex, amsmath, graphicx}
\title{Quantifying Educational Competition: A Game-Theoretic Model with Policy Implications}
\author{Siyuan He\thanks{Department of Economics, University College London}}
\date{December 2024}

\begin{document}

\maketitle

\begin{abstract}
    The competitive pressures in China's primary and secondary education system have persisted despite decades of policy interventions aimed at reducing academic burdens and alleviating parental anxiety. This paper develops a game-theoretic model to analyze the strategic interactions among families in this system, revealing how competition escalates into a socially irrational ``education arms race.'' Through equilibrium analysis and simulations, the study demonstrates the inherent trade-offs between education equity and social welfare, alongside the policy failures arising from biased social cognition. The model is further extended using Spence's signaling framework to explore the inefficiencies of the current system and propose policy solutions that address these issues. 
\end{abstract}

\section{Introduction}

\paragraph{Background: Education Pressure and Policy Challenges in China.} Chinese primary and secondary school students have long faced heavy academic pressure, driven by external demands or internal motivation to compete in a brutal ``education competition.'' A survey shows 75\% of students sleep less than 9 hours daily, and 45.2\% have less than 2 hours of free time (Li and Sun, 2004). Over the past 70 years (1949–2024), China has issued more than 10 policies to reduce this pressure, ranging from controlling class time and homework to regulating admissions, examinations, and evaluations (Zhang and Wan, 2018; Ge and Zhang, 2018). However, these measures have failed to effectively reduce students' class burden or alleviate parental anxiety (Wang and Liu, 2018; Liu and Dong, 2022; Xiang, 2019; Jia et al., 2021; Zhang and Wan, 2018; Zhu and Zhu, 2002).

\paragraph{Research Objectives.} In order to further understand this problem, this essay uses game theory to model the strategic interaction and social outcomes of different families within the education system. First, a game theory-based model is set up to represent the strategic interactions of students in China's education system. Then the equilibrium of this model is analyzed and the problem of competition escalation and collective irrationality is shown. A simulation based on the set of the model is then performed to discuss several previous policies to demonstrate the value dilemma: the trade-off between education equity and social welfare. Finally, the model is combined with Spence's (1973) signaling game theory to analyze the irrationality in the system and to propose policy solutions.

\section{Model Construction}

\subsection{Locate the social dilemma}

This essay aims to model overly competitive problems in China's education system: high-quality education resources are scarce, and exam results dominate competition. The reason the burden of learning cannot be eliminated is that it is a necessity for everyone in the society.

\paragraph{Social game characteristics of problems:} (1) Collective adherence to burden reduction rules increases overall welfare, but individual disobey brings individual benefits. (2) Strategic interaction between families causes loss of utility for everyone.

\subsection{Model Set-up}

We assume a student's score $S_i$ is determined by study time $t_i$ and student's aptitude $\gamma_i$. For simplicity, we omit any random exogenous factor (e.g. exam difficulty, student's performance, etc.) in this function.

\begin{equation}
    S_i = \gamma_i \cdot t_i
\end{equation}

Therefore, we have the average exam score of all students in the society $\bar S$, where $F(S_i)$ represents the cumulative distribution function of $S_i$.

\begin{equation} 
    \bar S = \int^{\infty}_0 (\gamma_i\cdot t_i) dF(S_i)
\end{equation}

Based on $\bar S$, we can formulate a score threshold $S_{cut}$ which determines whether a student is success in this competition:

\begin{equation}
    S_{cut} = \bar S + k\cdot \sigma_S
\end{equation}

The $\sigma_S$ is the standard deviation of scores of all families, and the coefficient $k$ indicates the severity of the threshold (for example, the 5\% percentile corresponds to $k \approx 1.645$). Therefore, this $S_{cut}$ defines a positional good: only when a student's exam score higher than $S_{cut}$ can they get the access of high-quality educational resource. Thus, based on the relative position of $S_i$ and $S_{cut}$, we can try to write down a family's utility function:

\begin{equation}
    u_i = \begin {cases}\log(2+S_i-S_{cut})-C_i, & \text{if } S_i \geq S_{cut} \\
    -C_i, & \text{if } S_i < S_{cut}\end {cases}
\end{equation}

This utility function can be easily interpreted as: if the student is successful in the competition, then the family get the payoff to hedge the cost of learning. But if it is not, then the family is paid nothing and they purely bear the loss of study cost for their kid. Also notice that here, we assume each family has only one child, and the family inevitably support their student to get academic success. $C_i$ is the cost of supporting student learning. Similar as $S_i$, it is also a function of $t_i$ (multiplies by time cost coefficient $P_i$:

\begin{equation}
    C_i = P_i \cdot t_i
\end{equation}  

If the expected $S_i < S_{cut}$ (that is, a family found even with effort, the threshold cannot be reached), the family will choose $t_i^*=0$, i.e. totally give up academic competition. Because in this case, the utility function is strictly decreasing with respect to $t_i$. However, in reality of China, constrained by traditional cultures that values the studying and conformity effect, most families will not choose to give up on learning (Xiang, 2019, p.76). Thus we assume most families expect $S_i \geq S_{cut}$ and apply this utility function:

\begin{equation}
    u_i = \log(2+S_i-S_{cut})-C_i
\end{equation}

Later we will analyze how this ``bounded rationality'' could be a root cause of China's ``educational arm race''. 

\section{Equilibrium Analysis}

\subsection{Check for concavity}

We let $x = 2+S_i-S_{cut}$. Thus $u_i = \log(x)-P\cdot t_i$. By definition, $x>0$. We need to take second-order condition of $u_i$ to check its concavity:

\begin{equation}
    \frac{\partial u_i}{\partial t_i} = \frac{\partial \log(x)}{\partial t_i}-P
\end{equation}

Solve to get:

\begin{equation}
    \frac{\partial u_i}{\partial t_i} = \frac{\gamma_i / 2}{x}-P
\end{equation}

Then we take second-order derivative of it:

\begin{equation}
    \frac{\partial^2 u_i}{\partial t_i^2} = -\frac{\gamma^2}{4x^2} < 0
\end{equation}

Therefore, the second-order derivative is less than zero, and we can confirm that the utility function $u_i$ is concave with respective to $t_i$. The only critical point found for this function is its global maximum.

\subsection{Two-family analysis}

For simplicity, we assume the society only contains two families, and the $S_{cut}$ is their average, $\frac{S_i + S_2}{2}$. In this scenario:

\begin{equation}
    u_i = \begin{cases} \log(2 + \gamma_i \cdot t_i- \frac{\gamma_i\cdot t_i + \gamma_j\cdot t_j}{2})-P\cdot t_i &\text{if } S_i\geq S_{cut} \\ -P\cdot t_i &\text{if }S_i < S_{cut} \end{cases}
\end{equation}

To analysis the best response of each family, we should begin with taking derivative to $u_i$ with respect to $t_i$:

\begin{equation}
    \frac{\partial u_i}{\partial t_i} = \frac{\frac{\gamma_i}{2}}{2+\gamma_i\cdot t_i - \frac{\gamma_i\cdot t_i + \gamma_j\cdot t_j}{2}} - P
\end{equation}

To maximize $u_i$, we need to find family i's best response $t_i^*$:

\begin{equation}
    t_i^* = \arg\max_{t_i} u_i \Longrightarrow \frac{\partial u_i}{\partial t_i} = 0
\end{equation}

Solve to get:

\begin{equation}
    t_i^* = \frac{1}{P} + \frac{\gamma_j\cdot t_j-4}{\gamma_i}
\end{equation}

This best response function tells us that the better the talent $\gamma_i$, the less the optimal learning time $t_i$. And the greater the learning time cost $P$, the less the optimal learning time $t_i$.

\paragraph{Strategic interaction dynamic.} We can clearly see the strategic interaction of family $i$ with other families in this $t_i^*$. When family $j$ increases their study time $t_j$, family $i$ needs to increase their own $t_i$ in response. And if family $j$ has a higher aptitude $\gamma_j$, the impact of its increased study time is amplified in family $i$'s best response formula. On the contrary, families with lower aptitude must invest more time to catch up with their competitors, further exacerbating their learning costs. This dynamic essentially creates a situation of an ``education arms race'': even if a family's investment cannot directly improve its relative ranking, it must continuously increase its study time to avoid falling behind.

\paragraph{Marginal utility diminishing.} Although increasing $t_i$ can improve scores, due to the concavity of $\log(2 +\cdot)$, the marginal returns of score growth become increasingly lower, while the cost $P \cdot t_i$ increases at a linear rate. This weakens the utility $u_i$ of the families. Finally, the learning time of all families tends to increase, but the utility $u_i$ may tend to decrease, as the cost gradually exceeds the benefits.

\subsection{Why the burden reduction policies are ineffective?}

Based on the given families' utility function, we then analyze the effect of whether families comply with the burden reduction policy in a simple strategic game (Zhu and Zhu, 2002). We Assume that there are only 2 families in the society, of which the students in family 1 are slightly more gifted than those in family 2, $\gamma_1=5, \gamma_2 = 4$. All parameters for setting up this game are shown in the Table 1.

\begin{table}[ht] %%参数： h:放在此处 t:放在顶端 b:放在底端 p:在本页
	\renewcommand\arraystretch{1.2}
	\centering  % 显示位置为中间
	\begin{tabular}{p{45pt}l|ll|l} %第一列设置宽度为45pt 全为左对齐 没有分割线
		\hline \hline  % 表格的横线
		%\toprule % 顶部线
		Parameters & & Definitions & Value & Type \\%[3pt]只改一行   
		\hline  % 表格的横线
		%\midrule % 中部线
            \multirow{2}[2]{*}{$\gamma_i$}& $\gamma_1$ & Student 1's Aptitude & 5 &  Fixed \\
		& $\gamma_2$ & Student 2's Aptitude   & 4 &  Fixed \\
		%\midrule
		$P$ & & Time Cost Coefficient & 0.5 &  {Fixed, Both} \\
            $t_{obey}$ & & Obey Policy Study Time & 2 &  {Fixed, Both} \\
		\multirow{2}[2]{*}{$t_i^*$} & $t_1^*$ & Family 1's Best Response & $\frac{4t_2+6}{5}$ &  Calculated \\
		& $t_2^*$ & Family 2's Best Response  & $\frac{5t_1+4}{4}$ &  Calculated \\
		%\bottomrule % 底部线
		\hline\hline  % 表格的横线
	\end{tabular}
        \caption{2*2 Strategic Game Setup.}
\end{table}

To start the game, we can first calculate the utility outcomes of family 1 in case of obeying the policy:

\begin{equation}
    u_1(obey) =\begin{cases}  \log (2+10-\frac{10+8}{2})-1  \approx 0.1 & \text{if family 2 obey}\\
-1 & \text{if family 2 disobey}  \end{cases}
\end{equation}

\paragraph{Computing details.} First, if both sides obey, the payoff of family 1 is calculated according to the study time of both sides is $t_{obey}=2$. And if family 2 disobey, we assume this family choose their best response $t_2^*$ based on family 1's choice. Here, for example, when $t_1=2$ (Because obeying the policy), family 2's best response is $t_2^*=\frac{14}{4}$. This give us their scores $S_1=2\times5=10$ and $S_2=\frac{14}{4} \times 4=14$, so family 1 loses the competition and only gets the utility of $-C_i$, i.e. $-0.5 \times 2= -1$. The following calculation methods are consistent, as long as either of the two families obey (i.e. $t_i=2$) and the other family disobey, then the utility of the other family is its best response utility calculated based on this $t_i$. We finally get the other payoff functions such as $u_1(disobey)$, $u_2(obey)$ and $u_2(disobey)$.

\begin{equation}
    u_1(disobey) =\begin{cases}  \log (2+14-\frac{14+8}{2})-\frac{7}{5}  \approx 0.21 & \text{if family 2 obey}\\
    \log5-\frac{4t_2+6}{10} & \text{if family 2 disobey}  \end{cases}
\end{equation}

\begin{equation}
    u_2(obey) =\begin{cases}  -1 & \text{if family 1 obey}\\
-1 & \text{if family 1 disobey}  \end{cases}
\end{equation}

\begin{equation}
    u_2(disobey) =\begin{cases} \log(2+14-\frac{10+14}{2})-\frac{7}{4} \approx -0.36 & \text{if family 1 obey}\\
    \log 4 - \frac{5t_1+4}{8} & \text{if family 1 disobey}  \end{cases}
\end{equation}

\begin{table}[ht]
    \centering
    \begin{game}{2}{2}
        & $\text{Family 1 Obey}$  & $\text{Family 1 Disobey}$\\
        $\text{Family 2 Obey}$ & 
        $(0.1,-1)$ & 
        $(0.21,-1)$\\
        $\text{Family 2 Disobey}$ & 
        $(-1,-0.36)$ & 
        $(\log 5 - \frac{4t_2+6}{10}, \log 4 - \frac{5t_1+4}{8})$
    \end{game} \\
    \caption{Strategic profile for two families with different $\gamma$.}
\end{table}

\paragraph{Analyzing the Strategic profile.} Family 1 and 2 tend to disobey because moving from an obey to a disobey state invariably increases their expected utility. Among them, for the less talented family 2, disobeying the ``burden reduction'' policy is the only chance to beat family 1 and win the competition. For family 1, choosing to disobey guarantees their winning of the competition. However, if they choose obey, there is a certain probability that they will fail and be surpassed by people with lower talents.

\paragraph{Unstable Nash equilibrium.} When both families choose to disobey, there is an unstable Nash equilibrium, which reflects a deteriorating dependence on each other's strategies: the increase in learning time for either side will reduce the other's payoff. We'll try to visualize this competitive escalation system later.

\paragraph{``Variance'' reduces social welfare.} When we modify student 2's aptitude to $\gamma_2 = 5$ and replay the game, a first-best appears in Table 3 (both families choose to obey). This game now has the same structure with the famous ``prisoner's dilemma'' if the we see ``disobey'' as ``betray''. Compared with Table 2, we found that introducing variance in $\gamma_i$ moves first-best from $(Obey, Obey)$ to $(Disobey, Obey)$ and slightly weakens the social utility of first-best from $-0.6$ to $-0.79$. 

\begin{table}[ht]
    \centering
    \begin{game}{2}{2}
        & $\text{Family 1 Obey}$  & $\text{Family 1 Disobey}$ \\
        $\text{Family 2 Obey}$ & $(-0.3,-0.3)$ & $(0.01,-1)$ \\
        $\text{Family 2 Disobey}$ & $(-1,0.01)$ & $(\log 4 - \frac{5t_2+6}{10}, \log 4 - \frac{5t_1+6}{10})$
    \end{game} 
    \caption{Strategic profile for two families with same $\gamma=5$.}
\end{table}

\subsection{Generalize the model: social simulation}

In two-family set, we assume the $S_{cut}$ is the average score of all students, namely $k=0$. This is a special case. In a more general way, 

\begin{equation}
    u_i = \begin {cases}\log(2+S_i-S_{cut})-C_i, & \text{if } S_i \geq S_{cut} \\
    -C_i, & \text{if } S_i < S_{cut}\end {cases}
\end{equation}

Similarly, we still focus on the first half of utility function because it represents the situation of most families. We take partial derivative to it with respect to $t_i$, and solve for $t_i^*$ which represents a family's choice in society (we assume that in a large society, the influence of $t_i$ with respect to $S_{cut}$ is negligible):

\begin{equation}
    t_i^* = \arg\max_{t_i} u_i \Longrightarrow \frac{\partial u_i}{\partial t_i} = 0
\end{equation}

And we finally get a new function of $t_i^*$. 

\begin{equation}
    t_i^* = \frac{1}{P} + \frac{S_{cut}-2}{\gamma_i}
\end{equation}

Similar as before, the higher $S_{cut}$ will lead to higher study time. The lower time cost $P$ leads to higher study time. Higher $\gamma_i$ will significantly scale down the affect of $S_{cut}$.

\subsection{From marginal utility diminishing: exhaustion caused by the arms race}

\paragraph{Maximum non-competitive learning time.} In the non-competitive case, every family has an upper boundary for study time. Due to the concavity of $\log(2 +\cdot)$, beyond this point, even if $S_{cut} = 0$, increasing $t_i$ could cause $u_i < 0$ and $u_i$ to decrease. This is the rational stopping point for learning time. Continuing may lead to visual impairment, excessive psychological pressure, and other problems.

\paragraph{Exceeding the maximum.} In reality, limited rationality caused by sunk costs and cultural norms often prevents families from quitting once they start. Similarly, this model tells us is that competition could push study time beyond the upper boundary: an increase in $S_{cut}$ may cause learners to further increase their study time. And the added learning time will cause the $u_i$ to decrease. To exemplify this, the Figure 1 shows the change in utility function of a same family when $S_{cut}$ increases from 0 to 3. The Table 4 lists all the parameters used to produce Figure 1.

\begin{figure}[ht]
    \centering
    \includegraphics[width=0.9\linewidth]{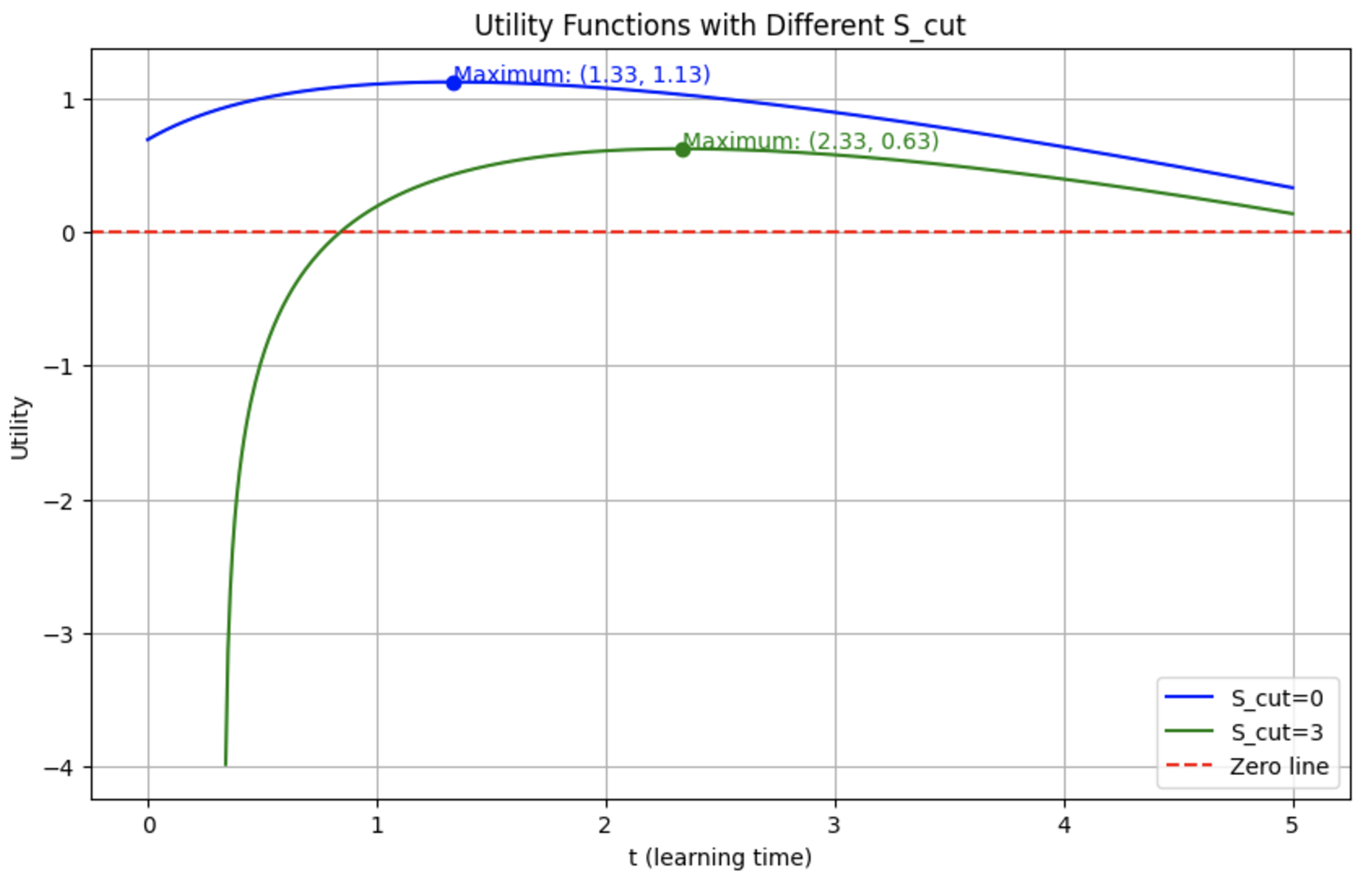}
    \caption{The difference in $S_{cut}$ affects same student's best response.}
\end{figure}

\begin{table}[ht] %%参数： h:放在此处 t:放在顶端 b:放在底端 p:在本页
	\renewcommand\arraystretch{1.2}
	\centering  % 显示位置为中间
	\begin{tabular}{p{55pt}l|ll} %第一列设置宽度为45pt 全为左对齐 没有分割线
		\hline \hline  % 表格的横线
		%\toprule % 顶部线
		Parameters && Definitions & Value \\%[3pt]只改一行   
		\hline  % 表格的横线
		%\midrule % 中部线
		$\gamma_i$ & & Student $i$'s Aptitude & 3 \\
		$P$ & & Time Cost Coefficient & 0.5 \\
		\multirow{2}[2]{*}{$S_{cut}$} & $S_{cut}^1$ & Set lowest threshold & 0 \\
		& $S_{cut}^2$ & Set the threshold to 3 & 3 \\
		%\bottomrule % 底部线
		\hline\hline  % 表格的横线
	\end{tabular}
        \caption{Figure 1's Parameters.}
\end{table}

\paragraph{Insights from Figure 1.} We discover that when $S_{cut} = 0$, a student with an aptitude of 3 only needs to study 1.33 hours per day to get their maximum utility of 1.13. When $S_{cut}$ is increased to 3, the same student would need to study 2.33 hours per day to achieve a maximum utility of 0.64. In other words, after $S_{cut}$ increased from 0 to 3, their learning time is increased by 75\%, while the utility is reduced by 43\%.

\paragraph{Relation between $S_{cut}$ and $t_i^*$.} $S_{cut}$ and individual best response $t_i^*$ constructs a mutually reinforcement structure: higher $S_{cut}$ will bring higher $t_i^*$. And because everyone adjusts learning time to achieve new $t_i^*$, a even higher $S_{cut}$ is again produced. And meanwhile, everyone's $u_i^*$ decreases.

\begin{equation}
    S_{\text{cut}} \uparrow \quad \implies t_i^* \uparrow \quad \implies \quad \bar{t} \uparrow \quad \implies \quad S_{\text{cut}} \uparrow
\end{equation}

When $t_i$ exceeds the rational upper limit of $u_i$ (i.e. $u_i < 0$ and $\frac{\partial u_i}{\partial t_i}\leq0$), the player $i$'s utility drops to a negative value. But because $S_{cut}$ is determined by the level of effort of others, individual rationality becomes ``one cannot stop trying even if rational decisions are ineffective''.

\section{Policy suggestion and evaluation}

Ideally, We want an education system to cover both aspects: to maximize the social welfare (utilitarian utility) and maximize the educational equity. In common sense, these two concepts are packed together: a more equitable and just education system maximizes the welfare of society. However, it is not the case in China's educational system. Next we will examine how our model represents this trade-off and evaluate some corresponding policies.

\subsection{Higher variance leading to lower social utility}

Inspired by two games represented by Table 2 and Table 3 in (3.3), we can intuitively understand that the increase in variance of students causes worse competition. Here we use simulation to exemplify this further. Figure 2 shows the increase in score distribution variance $\sigma_S$ exacerbates the competition by rising the $S_{cut}$. And a representative education policy of this is China's College Entrance Examination (``CEE'').

\begin{figure}[ht]
    \centering
    \includegraphics[width=0.9\linewidth]{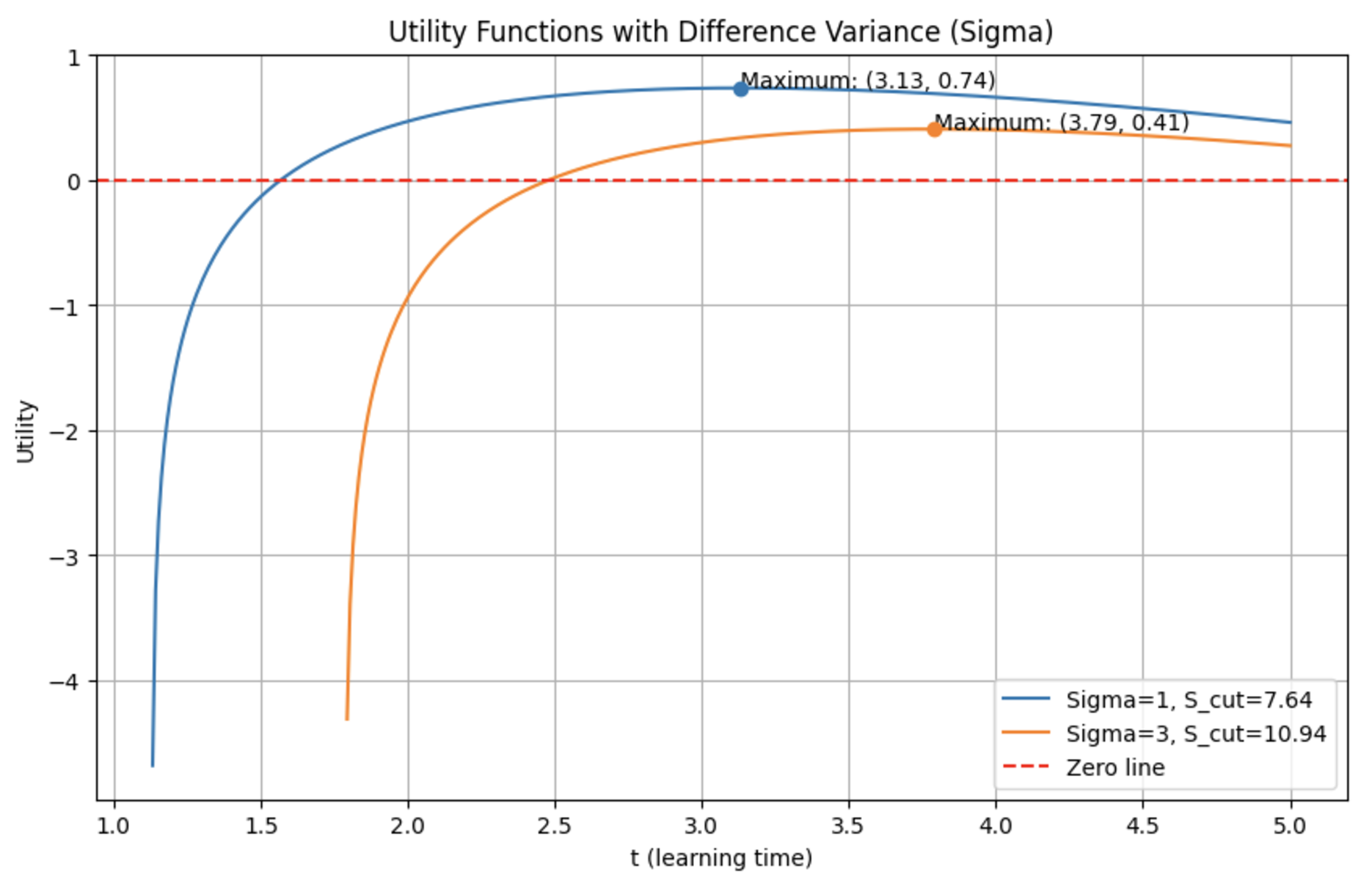}
    \caption{The difference in $\sigma_S$ affects same student's utility.}
\end{figure}

\begin{table}[ht] %%参数： h:放在此处 t:放在顶端 b:放在底端 p:在本页
    \renewcommand\arraystretch{1.2}
    \centering  % 显示位置为中间
	\begin{tabular}{p{45pt}l|ll|l} %第一列设置宽度为45pt 全为左对齐 没有分割线
		\hline \hline  % 表格的横线
		%\toprule % 顶部线
		Parameters & & Definitions & Value & Type \\%[3pt]只改一行   
		\hline  % 表格的横线
            $\gamma_i$ & & Student $i$'s Aptitude & 5 & Fixed  \\
            $\bar{\gamma}$ & & Students' Average Aptitude & 3 & Fixed \\
		$P$ & & Time Cost Coefficient & 0.5 &  Fixed \\
            $k$ & & $S_{cut}$ Coverage Coefficient & 1.645 & Fixed \\
            $\bar t$ && Average Study Time & 2 & Fixed \\
            \multirow{2}[2]{*}{$\sigma_S$} & $\sigma_S^1$ & The Lower Variance Setting & 1 &  Given, Fixed \\
		& $\sigma_S^2$ & The Higher Variance Setting & 3 &  Given, Fixed \\
		\multirow{2}[2]{*}{$S_{cut}$} & $S_{cut}^1$ & $S_{cut}$ When $\sigma_S=1$ & 7.64 &  Computed \\
		& $S_{cut}^2$ & $S_{cut}$ When $\sigma_S=3$  & 10.94 &  Computed \\
		\hline\hline  % 表格的横线
	\end{tabular}
    \caption{Figure 2's parameters}
\end{table}

\paragraph{Policy 1: ``CEE'' Policy.} CEE makes all students can go to the same examination room to fight for opportunities equally. However, this ``participate by all people'' feature makes it naturally a strong educational arms race. The high participant variance causes $S_{cut}$ continue to rise, and social welfare continue to decrease.

\subsection{Lower variance brings the less education equity}

\paragraph{Policy 2: 50-50 Diversion Policy.} Since 2018, China has issued a 50\%-50\% diversion policy to alleviate academic pressure of CEE (Anon, 2023). This means that only half of middle school students can take the traditional college entrance exam to take the academic path, and students who do not score in the top 50 percent of the high-school entrance exam must pursue vocational education after middle school. According to our model, it reduces the variance $\sigma_S$ of students by shrinking the population, thus reducing the $S_{cut}$, that is, reducing the academic competition in high school. However, this policy reduces educational equity by depriving half the students of the right to compete for higher education resources. For these ``abandoned'' students and families, the critical thing in implementing this policy is whether the utility of the subsidy to them is sufficient.

\section{Discussion: an optimal policy revealed by our model}

The derivation above has been based on the assumption that most families in the society will not choose other paths than education competition or withdraw from education competition in the halfway because of irrational factors such as cultural conventions, sunk costs, path dependence and herd effect (Xiang, 2019). It is clear that this irrationality is an essential factor leading to excessive competition in education, but we have not yet analyzed its causes. Combined with the signal game model proposed by Spence (1973), irrationality in decision making can be measured by our model, and a better policy to this problem will likely be revealed.

\subsection{Expand the model with Spence (1973)}

\paragraph{New Model Set-up.} One important claim Spence (1973) made was: ``individuals' goal is to maximize the difference between their expected future wages and signaling costs (i.e. education cost here)''. To illustrate this, we assume high-quality education will bring 2,000 pounds on average wage ($W_{high}$) for an individual, and without that an individual will only find jobs with 1,000 pounds wage on average ($W_{low}$). We also know that lower aptitude $\gamma_i$ leads to more learning time cost $C_i$.

\begin{equation}
    \gamma_i \downarrow \quad \implies \quad t_i^* \uparrow \quad \implies \quad C_i \uparrow
\end{equation}

Therefore, assuming that a child of Family 1 has a low talent of $\gamma_1$, and their time cost of learning ($C_i$) is 1500 pounds to achieve $S_{cut}$, so that they can reach a higher education level to get the job of 2,000 pounds salary. Otherwise they only get paid 1,000 pounds. Then their rational choice is not to start learning at all. Because if they choose not to go to school, the payoff will be $1000-0 = 1000$ (costless). And if he chooses to study and reaches $S_{cut}$, he will only have a payoff of $2000-1500 = 500$ pounds. Following the same logic, if Family 2's child has a higher aptitude $\gamma_2$, making them only cost $C_i = 800$ to achieve $S_{cut}$, then Family 2 will choose to invest the learning time to just exceed $S_{cut}$ to obtain the maximum difference of 1200 pounds ($2000-800$) as their payoff. As shown in Figure 3, people with higher ability get high salary, and others get low salary. A signaling equilibrium is achieved.

\begin{figure}[ht]
    \centering
    \includegraphics[width=0.9\linewidth]{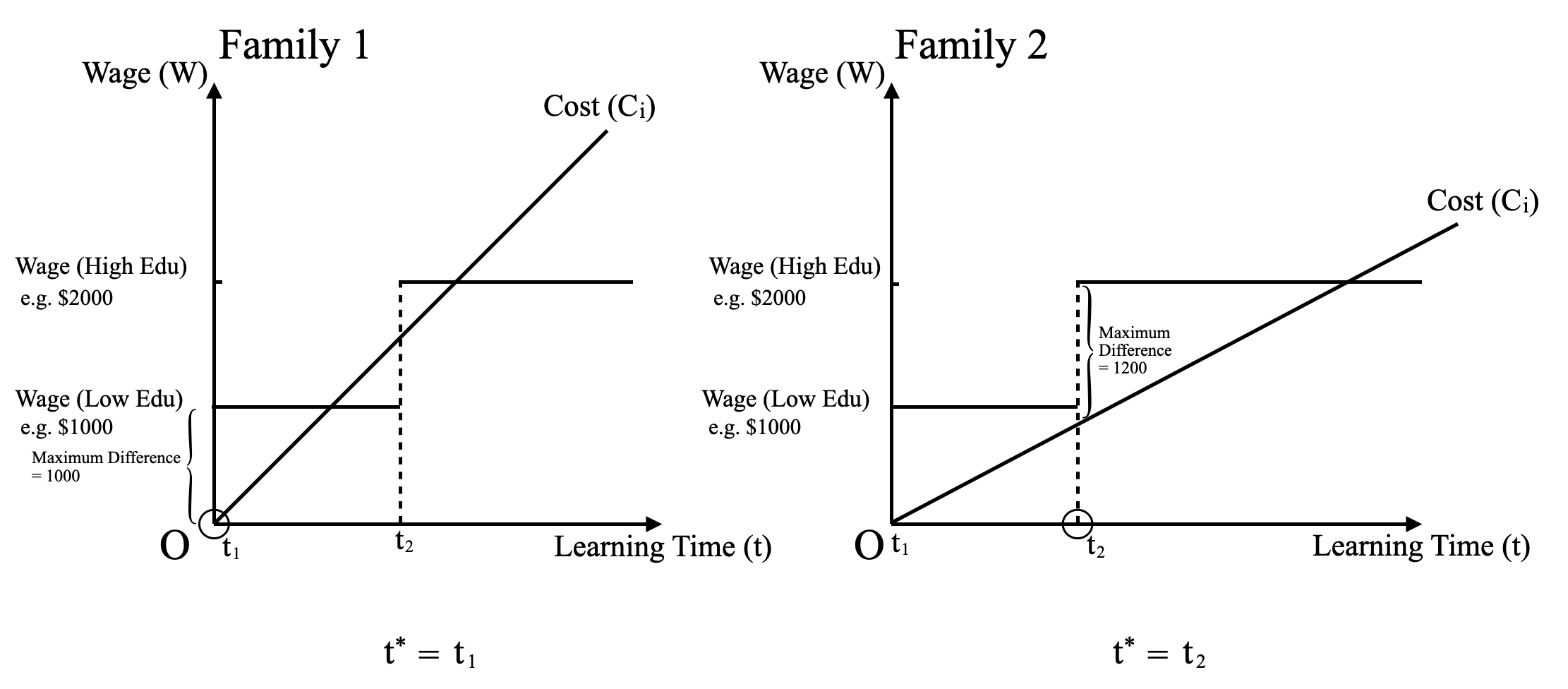}
    \caption{Difference in $\gamma_i$ let families behave differently in education decision-making.}
\end{figure}

\paragraph{Update the Model.} Equation (23) is the updated payoff function for families to include wage's impact on educational decision. The new added wage variables ($W_{low}$ and $W_{high}$) are essentially reverse signals to job seekers (students). And it allow families to plan their learning time more rationally. If the wage $W_{high}$ premium for higher education level is high in the market, students will tend to extend their study time to obtain a higher education level. If the premium is lower, students may choose to enter the labor market directly.

\begin{equation}
    u_i^{new} = \begin {cases} (W_{high} \cdot \beta) \times \log(2+S_i-S_{cut})-C_i, & \text{if } S_i \geq S_{cut} \\
    (W_{low})-C_i, & \text{if } S_i < S_{cut}\end {cases}
\end{equation}

\paragraph{Understanding the irrationality.} This expanded model prompts us to understand the irrational behavior of China's education system: the coefficient $\beta$ of $W_{high}$ reflects a cognitive bias, which is people may mentally magnify the extra benefit of $S_i > S_{cut}$ by more than 10 times (e.g. $\beta > 10$), Therefore, when making decisions, they believe that better academic rankings are everything sufficient of success, and do not give up studies no matter what.

\subsection{Proposed Policy Solution: Educational Decision Making to Avoid Irrational Factors}

Zhu and Zhu (2018) identified two hypotheses contributing to problems in China's education system: first, educational resources, especially higher education, are scarce; second, there is a intrinsic link between a student’s academic (or exam) performance and future welfare distribution. Weakening this relationship, both in reality and public perception, should lead to more rational educational decisions for families (i.e. the variance $\sigma_S$ of educational competition will be reduced). This would lower resource waste, increase probability of self-actualization, and improve social welfare. Based on this, below I outline a policy solution consisting of three main points.

\paragraph{1. Reinforcing Diverse Career Signal.} By strengthening more possibilities for career development (e.g. promoting other career development possibilities through mass media, etc.), and weakening irrational factors such as conformity effect (reducing $\beta$), people can see their real benefits, so that ``quitting education competition'' hopefully becomes a feasible choice for most families in reality.

\paragraph{2. Improving Diverse Career Supporting System.} Improve the diversified career development and promotion system at the practical level, for example, learn from the ``dual track system'' of the German education system. Allow students to explore more fields outside the academic system to have practical social recognition and support (i.e. increasing the subsidy mentioned in 4.2).

\paragraph{3. Reduce Study Fatigue by Improving Exam Design.} In order to reduce the burden on students who choose to participate in academic education, reduce the dependence of test score $S_i$ on study time $t_i$ (emphasizing aptitude, e.g. choosing questions that emphasize creativity rather than time spent memorizing).

\section{Conclusion}

This essay employs a game-theoretic model to analyze the ``education arm race'' problem in China's education system. We have discovered the strategic pattern that causes the policy failure of burden-reduction policy, the trade-off between education equity and social welfare,  the irrationality problem caused by social cognition distortion, and finally proposed seemingly feasible solution based on our analysis. 

\paragraph{The transferability of the model.} The proposed model can be extended beyond the context of education. For instance, it can be applied to analyze strategic interactions between employers and employees, as explored in Dewatripont, Jewitt, and Tirole (1999). More broadly, the model is adaptable to any context characterized by escalating competition that results in resource inefficiency. Examples include housing markets, allocation of medical resources, corporate competition for environmental policy incentives, and even the over-training phenomenon in competitive sports.

\paragraph{Limitations and future directions.} By sharing this model, this article hope to inspire further research in both theoretical refinements and empirical validations as it does not incorporate any real-world data at this stage. Also, further incorporating the dynamic decision-making process in competitive games, where families adjust their strategies based on evolving factors such as resource consumption, child scores, competitors' scores, and economic constraints, would require more variables and a formalized expectation threshold for families' decisions. Additionally, as the game actually features asymmetric information, future work could explore Bayesian Games to model families' belief updates and strategies under uncertainty. Finally, the uncertainty around achieving the education signal $S_{cut}$ – stemming from imperfect knowledge of aptitude $\gamma_i$ and competition –is only partially addressed here, leaving room for simulation-based insights on optimal effort strategies across varying population profiles.

\end{document}